\documentclass[aps,prd,onecolumn,showkeys,showpacs,amssymb]{revtex4}
\usepackage{epsfig}
\usepackage{graphics}
\usepackage{graphicx}
\usepackage{amsmath,amssymb}
\usepackage{amsfonts}
\usepackage{bm}
\usepackage[usenames,dvipsnames]{color}
\usepackage{amssymb}
\usepackage{psfrag}
\usepackage{times}
\usepackage{xcolor}
\usepackage[varg]{txfonts}
\usepackage{xspace}
\usepackage{float}
\usepackage{placeins} 
\usepackage{capt-of} 

\begin{document}
\title{Extended Gravity Description for the GW190814 Supermassive Neutron Star}
\author{A.V. Astashenok$^{1}$, S. Capozziello$^{2,3}$, S. D. Odintsov$^{4,5}$, V. K. Oikonomou$^{6,7}$,  }
\date{\today}

\affiliation{$^{1}$ Institute of Physics, Mathematics and IT, I.
Kant Baltic Federal University, Kaliningrad, 236041 Russia}

\affiliation{$^{2}$ Dipartimento di Fisica, ``E. Pancini''
Universit\`{a} ``Federico II'' di Napoli,
Compl. Univ. Monte S. Angelo Ed. G, Via Cinthia, I-80126
Napoli (Italy)}

\affiliation{$^{3}$ INFN Sez. di Napoli, Compl. Univ.
Monte S. Angelo Ed. G, Via Cinthia, I-80126 Napoli, Italy}

\affiliation{$^{4}$
Institute of Space Sciences (IEEC-CSIC) C. Can Magrans s/n, 08193 Barcelona, Spain}

\affiliation{$^{5}$
ICREA, Passeig Luis Companys,
23, 08010 Barcelona, Spain}

\affiliation{$^{6}$ Department of Physics,
Aristotle University of Thessaloniki,
Thessaloniki 54124 Greece}

\affiliation{$^7$ Int. Lab. Theor. Cosmology,
Tomsk State University of Control Systems and Radioelectronics (TUSUR),
634050 Tomsk, Russia.}

\begin{abstract}
Very recently a compact object with a mass in the range $2.50\div 2.67\,
M_{\odot}$ has been discovered via gravitational waves detection
of a compact binary coalescence. The mass of this object makes it among
the heaviest neutron star never detected or the lightest black hole ever
observed. Here we show that a neutron star with this
observed mass, can be consistently explained  with   the mass-radius relation obtained
 by  Extended Theories of Gravity. Furthermore,
 equations of state, consistent with LIGO observational constraints, are adopted. We  consider also the
influence of rotation
and show that masses of rotating neutron stars can exceed $2.6
M_\odot$ for some equations of state  compatible with  LIGO data.

\end{abstract}


\maketitle

\section{Introduction}
Compact astrophysical objects, such as Neutron Stars (NSs),  can be described by General Relativity
(GR) in the strong field regime. The structure of a NS is strictly
correlated with the equation of state (EoS), {\it i.e.} the
relation between pressure and density in its interior
\cite{Lattimer}. Given an EoS, a mass-radius $(M-\cal{R})$
relation  and a corresponding maximal mass can be derived, in
principle, for any NS. Furthermore, the knowledge of these
parameters provides significant information on the mechanism
responsible for  formation, stability and possible effects on  the
evolutionary history of NSs. For a detailed introduction to the
theory of relativistic stars, see for example \cite{psaltis}.

On the other hand, NSs are natural laboratories to test strong
gravitational field regimes that can hardly be reached in any
other part of the Universe and, so, their internal structure
cannot be easily reproduced because of the extreme conditions in
which it operates. Thus,   theoretical models can be formulated
where  a very large number of EoS candidates can be taken into
account. Starting from microscopic information, the task is  to
reproduce consistently the observed macroscopic parameters, and,
viceversa, from these parameters, to select and constrain reliable
classes of EoS. In this perspective,  astrophysical measurements
of  mass, radius and rotation, besides
 selecting  realistic EoS, give also
insight on the behavior of  matter in  extreme gravity regimes.

It is important to stress that GR gives  very strict limits for
the stability of compact objects made up of degenerate matter,
like NSs or white dwarfs. In particular, Chandrasekhar fixed a
theoretical upper bound of  ${\cal M}\sim 1.44 \,M_{\odot}$
\cite{Chandrasekhar} for non-rotating compact objects. For  masses
around this limit, gravity and degenerate matter achieve stable
configurations around a radius of ${\cal R}\sim 10$ Km. Beyond
this limit, considering also secondary effects which can improve
it, gravity cannot be stopped by degenerate matter pressure and
black holes originate.

From an observational point of view, the mass determination can
accurately be achieved only for NSs in binary systems.
Observations of these systems have found some NSs that could
violate this limit
\cite{Barziv:2001ad,Rawls:2011jw,Mullally:2009rr,Nice:2005fi,2010Natur.467.1081D,Bao}.
In particular, a very recent observation \cite{Ligo} detected a
compact object in the mass range $2.50 \div 2.67\,M_{\odot}$ via
gravitational waves. It  could represent the most massive NS ever
observed. This result is well beyond the Chandrasekhar limit  so
it cannot be in agreement with the standard theory also stretching
parameters and processes  in the  GR context. The way out to these
shortcomings could be finding some exotic EoS capable of
stabilizing the stellar structure by some form of degenerate,
strange matter, or considering  alternative gravity where
Chandrasekhar limit can be relaxed or improved. On the other hand,
the massive object could be a very light black hole, but also in
this case there are difficulties in  explaining it by the standard
theory.

In the  research line of modified gravity alternative
description, Extended Theories of Gravity (ETGs)
\cite{Capozziello:2011et} could play a prominent role in
explaining consistently the problem at hand. Such theories are
straightforward extensions of GR. Specifically, GR is a particular
case of a large class of theories which proved to be particularly
useful in the IR limit of cosmology (see \cite{dagostino} for a
recent review). Detailed studies of anomalous astrophysical
compact objects in the framework of ETGs have been already
performed in some  previous works under the hypothesis that very
massive NSs could be materialized by gravitational (geometric)
effects
\cite{capquark,Astashenok:2013vza,Astashenok:2014pua,Astashenok:2014nua}.
 In particular   $f(R)$ gravity, i.e.  a
class of Lagrangians considering   a generic function of the Ricci
curvature scalar, has been investigated. Clearly, for $f(R)\to R$ the standard GR is restored.

$f(R)$ gravity has been used to solve and explain theoretically a
large number of astrophysical and cosmological issues, {\it i.e.}
the cosmic acceleration \cite{Perlmutter:1998np,Riess:1998cb,
Riess:2004nr,Spergel:2006hy,Schimd:2006pa,McDonald:2004eu}, the
inflationary paradigm,  the dark matter,
 the dark energy, and some stellar structures \cite{Capozziello:2002rd, Capozziello:2003tk,
Nojiri:2003ft, Carroll:2003wy, Olmo:2011uz, Nojiri:2010wj,
Capozziello:2010zz, delaCruzDombriz:2012xy}.

For the astrophysical GW190814 event, the primary objective  is to
obtain the $\mathcal{M-R}$ relation of  NSs allowing to derive the
maximum mass value. This result should be achieved considering
realistic EoS.

In this paper, we want to demonstrate that,  measurements reported
by the LIGO collaboration \cite{Ligo,TheLIGOScientific:2017qsa}
for the GW190814 event, can be theoretically framed in the context
of ETGs, if the $\mathcal{M-R}$ relation is  obtained by a system
of modified Tolman-Oppenheimer-Volkoff (TOV)  equations
\cite{Feola,Capozziello2016}. Clearly, in the limit $f(R)\to R$,
the standard TOV system is recovered \cite{Oppenheimer:1939ne}.

The paper is organized as follows. In Sect. II we derive the TOV
equations for $f(R)$ gravity. Specific $f(R)$ models are also
discussed. Then,  in Sect. III, we consider  rotating stars in the
framework of $f(R)$ gravity.  Sect. IV is devoted to the numerical
results for  static and rotating cases.  Conclusions are drawn in
Sec. V.

\section{The Tolman-Oppenheimer-Volkoff equations in $f(R)$ gravity}

Let us start from the  $f(R)$  action given by

\begin{equation}\label{action}
{\cal A}=\frac{c^4}{16\pi G}\int d^4x \sqrt{-g}\left[f(R) + {\cal L}_{{\rm
matter}}\right]\,,
\end{equation}
where $g$ is the determinant of the metric $g_{\mu\nu}$ and ${\cal L}_{\rm
matter}$ is the standard perfect fluid matter Lagrangian. The
variation of (\ref{action}) with  respect to  $g_{\mu\nu}$ gives
the field equations \cite{Capozziello:2011et,Nojiri:2010wj,Mauro,Capozziello:2010zz}:

\begin{equation}
\frac{df(R)}{d R}R_{\mu\nu}-\frac{1}{2}f(R) g_{\mu\nu}-\left[\nabla_{\mu} \nabla_{\nu} - g_{\mu\nu} \Box\right]\frac{df(R)}{dR}=\frac{8\pi G}{c^{4}} T_{\mu \nu },
\label{field_eq}
\end{equation}
where $\displaystyle{T_{\mu\nu}= \frac{-2}{\sqrt{-g}}\frac{\delta\left(\sqrt{-g}{\cal L}_m\right)}{\delta g^{\mu\nu}}}$ is the energy momentum tensor of matter.
Here we  adopt the signature  $\left(+,-,-,-\right)$.
The metric for systems with spherical symmetry has the usual form

\begin{equation}
    ds^2= e^{2\psi}c^2 dt^2 -e^{2\lambda}dr^2 -r^2 (d\theta^2 +\sin^2\theta d\phi^2),
    \label{metric}
\end{equation}
where $\psi$ and $\lambda$ are functions depending only on the radial coordinate $r$.
Within the stellar structure, matter is described  as a perfect fluid,  whose  energy-momentum tensor is $T_{\mu\nu}=\mbox{diag}(e^{2\psi}\rho
c^{2}, e^{2\lambda}p, r^2 p, r^{2}p\sin^{2}\theta)$. Here $\rho$ is the
matter density and $p$ is the pressure \cite{weinberg}.
The equations for the stellar configuration are obtained adding the condition of hydrostatic equilibrium which can be derived from the contracted Bianchi identities

\begin{equation}
\nabla^{\mu}T_{\mu\nu}=0\,,
\label{bianchi}
\end{equation}
that give  the  Euler conservation equation
\begin{equation}\label{hydro}
    \frac{dp}{dr}=-(\rho
    +p)\frac{d\psi}{dr}\,.
\end{equation}
From the metric \eqref{metric} and the field equations (\ref{field_eq}), it is possible
to derive the equations for the functions $\lambda$ and $w$ in the form \cite{capquark}

\begin{eqnarray}
\label{dlambda_dr}
\frac{d\lambda}{dr}&=&\frac{ e^{2 \lambda }[r^2(16 \pi \rho + f(R))-f'(R)(r^2 R+2)]+2R_{r}^2 f'''(R)r^2+2r f''(R)[r R_{r,r} +2R_{r}]+2 f'(R)}{2r \left[2 f'(R)+r R_{r} f''(R)\right]},
\end{eqnarray}
\begin{eqnarray}\label{psi1}
\frac{d\psi}{dr}&=&\frac{ e^{2 \lambda }[r^2(16 \pi p -f(R))+ f'(R)(r^2 R+2)]-2(2rf''(R)R_{r}+ f'(R))}{2r \left[2 f'(R)+r R_{r} f''(R)\right]},
\end{eqnarray}
respectively. In both Eqs. (\ref{dlambda_dr}) and (\ref{psi1}), the prime denotes a derivative
with respect to   the Ricci scalar $R(r)$.

The above equations are the modified TOV equations that, for
$f(R)=R$, reduce to the standard  TOV equations of GR
\cite{rezzollazan,landaufluid}. It is important to stress that, in
$f(R)$ gravity, the Ricci scalar is a dynamical variable and then
we need a further equation to solve the system of Eqs.
(\ref{hydro}),  (\ref{dlambda_dr}) and (\ref{psi1}). The
corresponding equation  takes the form
\begin{equation}\label{TOVR}
\frac{d^2R}{dr^2}=R_{r}\left(\lambda_{r}+\frac{1}{r}\right)+\frac{f'(R)}{f''(R)}\left[\frac{1}{r}\left(3\psi_{r}-\lambda_{r}+\frac{2}{r}\right)-
e^{2 \lambda }\left(\frac{R}{2} + \frac{2}{r^2}\right)\right]-
\frac{R_{r}^2f'''(R)}{f''(R)},
\end{equation}
which can be derived from the trace of Eqs.\eqref{field_eq}
inserting the metric \eqref{metric}.

Let us now consider two physically motivated functional forms of
the $f(R)$ function and derive the TOV equations for these cases.
The aim is to demonstrate that minimal modifications with respect
to GR can give relevant results capable of explaining consistently
the problem of having supermassive NSs, without the need for a
stiff EoS. In fact, our description can incorporate even more
massive NSs, which may eventually will not be able to be described
by standard GR, even with stiff EoS being used.

\subsection{The  $f(R)=R+\alpha R^2$ model}
We consider here the specific form of $f(R)$:
\begin{equation}
f(R)=R+\alpha R^2,
\label{fr_form_quadratic}
\end{equation}
where $\alpha$ is the coupling parameter of the quadratic
curvature correction. This model is specially suitable to account
for cosmological inflation, where higher-order curvature terms
naturally lead to cosmic accelerated expansion. The quadratic term
emerges in strong gravity regimes and as an effective contribution in quantum field theory on curved spacetime \cite{Birrell}.  However, at Solar System scales
and, more in general, in the weak field regime, the linear term
predominates.

It is worth noticing  that in the interior of a NS, the physical
conditions quantified by the energy and pressure, could be
analogous to those during the early Universe
\cite{Astashenok:2014pua}. Due to this feature, the model
\eqref{fr_form_quadratic} is particularly suitable for our
considerations. Specifically, Eqs. (\ref{dlambda_dr}),
(\ref{psi1}) and (\ref{TOVR}) take the explicit form:
\begin{eqnarray}
\label{dlambda_dr alfa}
\frac{d\lambda}{dr}&=&\frac{ e^{2 \lambda }[16 \pi r^2 \rho - 2-\alpha R(r^2 R+4)]+4\alpha(r^2 R_{r,r} +2 r R_{r}+R)+2}{4r \left[1+\alpha(2R+r R_{r})\right]},
\end{eqnarray}
\begin{eqnarray}\label{psi1 alfa}
\frac{d\psi}{dr}&=&\frac{ e^{2 \lambda }[16 \pi r^2 p + 2+\alpha R(r^2 R+4)]-4\alpha(2rR_{r}+R)-2}{4r \left[1+\alpha(2R+r R_{r})\right]},
\end{eqnarray}
\begin{equation}\label{TOVR alfa}
\frac{d^2R}{dr^2}=R_{r}\left(\lambda_{r}+\frac{1}{r}\right)+\frac{1+2\alpha R}{2\alpha}\left[\frac{1}{r}\left(3\psi_{r}-\lambda_{r}+\frac{2}{r}\right)- e^{2 \lambda }\left(\frac{R}{2} + \frac{2}{r^2}\right)\right]\,.
\end{equation}
Clearly, GR is restored for $\alpha = 0$.

\subsection{The $f(R)=R^{1+\varepsilon}$ model}
\label{sect_power}

Another interesting class of models are the power law models
$f(R)\sim R^n$ with $n\in \mathbb{R}$. As shown in \cite{Stabile},
these models are related to the existence of Noether symmetries.
For $n=1$, the Noether symmetry gives the standard Schwarzschild
radius as a conserved quantity. We can assume the form
\begin{eqnarray}\label{LOGe}
f(R)=R^{1+\varepsilon}\,,
 \end{eqnarray}
where $n=1+\varepsilon$, to study small deviation with respect to GR for
  $|\varepsilon| \ll 1$. In this limit,    it is possible to write
 a first-order Taylor expansion as

 \begin{eqnarray}\label{LOG}
R^{1+\varepsilon}&\simeq & R+\varepsilon R {\rm log}R +O (\varepsilon^2) ,
 \end{eqnarray}
 which is  interesting in order to define the
correct physical dimensions of the coupling constant and to
control the magnitude of the corrections with respect to the
standard Einstein gravity \footnote{It is important to stress that
also this kind of corrections emerges in one-loop regularization
and renormalization process in curved spacetime \cite{Birrell}.}.

A term in the Lagrangian of the form  (\ref{LOG}) has been widely
tested starting from Solar System up to cosmological scales.
Indeed, the value of the parameter $\varepsilon$ can
straightforwardly  relate a weak field curvature regime
$(\varepsilon \simeq 0)$ to a regime where strong curvature
effects start to become relevant $(\varepsilon \neq 0)$. In this
perspective,   $\varepsilon$ could be different from zero in NSs
and then probe deviations with respect to GR. The explicit forms
of Eq. (\ref{dlambda_dr}) and (\ref{psi1}) for the action
(\ref{LOG}) are:

\begin{eqnarray}\label{tovlambda_power}
\frac{d\lambda}{dr}=\frac{8 e^{2 \lambda} G \pi r R \rho}{c^2 [2 R (1 + \varepsilon + \varepsilon {\rm log}R) + r \varepsilon R']}
+\frac{ e^{2 \lambda} R^2 [r^2 R \varepsilon - 2 (1 + \varepsilon + \varepsilon {\rm log} R)]
+2 [R^2 (1 + \varepsilon + \varepsilon {\rm log} R) -
    r^2 \varepsilon R'^2 +
    r R \varepsilon (2 R' + r R'')]}{2 r R [2 R (1 + \varepsilon + \varepsilon {\rm log} R) +
       r \varepsilon R']}\,
\end{eqnarray}
and
\begin{eqnarray}\label{tovpsi_power}
\frac{d\psi}{dr}=\frac{8 e^{2 \lambda} G P \pi r R}{c^4 [2 R (1 + \varepsilon + \varepsilon {\rm log} R) + r \varepsilon R']}
-\frac{e^{2 \lambda}
   R [r^2 R \varepsilon - 2 (1 + \varepsilon + \varepsilon {\rm log} R)]
 +2 [R (1 + \varepsilon + \varepsilon {\rm log} R) +
    2 r \varepsilon R']}{2 r [2 R (1 + \varepsilon + \varepsilon {\rm log} R) + r \varepsilon R']},
\end{eqnarray}
while the equation for R is
\begin{eqnarray}\label{trace_power} \frac{d^2R}{dr^2}=\frac{R'^2}{R}+R'\left(\lambda'-\frac{2}{r}-w' \right)  -\frac{e^{2 \lambda} R [c^4 R ( 1-\varepsilon) + 8 G \pi (3 P - c^2 \rho) + c^4 R \varepsilon {\rm log} R]}{3 c^4 \varepsilon}.
\end{eqnarray}
Also in this case, GR is restored for $\varepsilon\to 0$. The
final aim of this mathematical apparatus is to investigate if
physical relations  of supermassive NSs, like the $\mathcal{M-R}$
diagram, can be realized by modified TOV systems according to the
values of parameters $\alpha$ and  $\varepsilon$. Before tackling
this task, let us discuss  also rotating NSs in the framework of
$f(R)$.

\section{Rotating neutron stars in $f(R)$ gravity}

Studying   spinning NSs is very important from a
theoretical point of view because realistic stellar structures are always rotating objects. NSs in binary systems, after
merging,  can produce black holes or supermassive fast rotating
NSs which then collapse into black holes \cite{FM}.
Parameters of post-merging gravitational wave signals are strongly
depending on angular momenta, masses and other secondary parameters of
NSs so then a multimessenger characterization of relativistic stellar objects could help also in selecting the  theory of gravity working in these systems.

Let us consider now a star rotating along the polar axis with angular
frequency $\Omega$. It is convenient to use metric in
quasi-isotropic coordinates namely
\begin{equation}\label{Metr}
ds^2 = e^{2\psi} c^2 dt^2 - e^{2\lambda}(dr^2 + r^2 d\theta^2) -
e^{2\mu} r^2 \sin^{2} \theta (d\phi - \omega dt)^2,
\end{equation}
where metric functions $\psi$, $\lambda$, $\mu$ and $\omega$
depend only on coordinates $r$ and $\theta$. It is worth  noticing that
$\lambda$, in this metric, does not reduce immediately to  $\lambda$ in the
previous section also  in the limit $\omega\rightarrow 0$.

 In GR, a $(3+1)$ formalism is usually
adopted for rotating stars (see for details  \cite{Alcubierre,Shapiro,Friedman}). In the case of $f(R)$ gravity, being this theory a straightforward extension of GR,  the same
formalism can be used without significant changes. Dropping
technical details, let us give the system of field equations
\begin{equation}\label{EQ_N}
f'(R)\Delta_{(3)}\psi + \frac{1}{2}\Delta_{(3)} f'(R)=4\pi
e^{2\lambda} (\epsilon+\sigma)-\frac{1}{2}e^{2\lambda}
(f'(R)R-f(R)) -
\end{equation}
$$
-f'(R)\partial \psi \partial (\psi+\mu)-\partial \psi
\partial f'(R) -\frac{1}{2}\partial \ln (\psi+\mu)
\partial f'(R) + f'(R) \frac{1}{2}{e^{2(\mu-\psi)} r^2 \sin^{2}\theta}(\partial
\omega)^{2},
$$

\begin{equation}\label{EQ_NB}
f'(R)\Delta_{(4)}(\psi+\mu) + \Delta_{(4)} f'(R) = 8\pi
e^{2\lambda} ( \sigma^{r}_{r} + \sigma^{\theta}_{\theta} ) -
e^{2\lambda} (f'(R)R-f(R)) -
\end{equation}
$$
-f'(R) (\partial  (\psi+\mu))^{2} - 2\partial (\psi+\mu)\partial
f_{R},
$$

\begin{equation}\label{EQ_NA}
f'(R)\Delta_{(2)}(\psi+\lambda)+\Delta_{(2)} f'(R) = 8\pi
e^{2\lambda} \sigma^{\phi}_{\phi} - \frac{1}{2}e^{2\lambda}
(f'(R)R-f) -
\end{equation}
$$
- f'(R) (\partial \psi)^{2} - \partial \psi
\partial f'(R) + \frac{3}{8}f'(R) {e^{2(\mu-\psi)} r^2 \sin^2 \theta
}{}(\partial\omega)^{2},
$$

\begin{equation}\label{EQ_omega}
f'(R)\Delta_{(5)}\omega = -  \frac{16\pi
e^{\psi+2(\lambda-\mu)}}{r^{2} \sin^{2} \theta}p_{\phi} +
\end{equation}
$$
+f''(R)\left[\partial R \partial \omega + 4\omega \partial \mu
\partial R +\frac{4\omega}{r}\left(\frac{\partial R}{\partial
r}+\frac{1}{r\tan\theta}\frac{\partial R}{\partial
\theta}\right)\right] -3 f'(R) \partial \mu \partial \omega +
f'(R)\partial \psi
\partial \omega.
$$
For any two given quantities $g_1$ and $g_2$, we define, for brevity,   the  notation
$$
\partial g_{1} \partial g_{2} \equiv \left(\frac{\partial g_{1}}{\partial
r}\frac{\partial g_{2}}{\partial r}+\frac{1}{r^2}\frac{\partial
g_{1}}{\partial \theta}\frac{\partial g_{2}}{\partial
\theta}\right).
$$
 $\Delta_{(n)}$ defines  the Laplace operators in Euclidean space
 including derivatives of radial
and polar coordinates:
$$
\Delta_{(n)}=\frac{1}{r^{n-1}}\frac{\partial}{\partial
r}\left(r^{n-1}\frac{\partial}{\partial r}\right)+\frac{1}{r^2
\sin^{n-2}\theta}\frac{\partial}{\partial
\theta}\left(\sin^{n-2}\theta \frac{\partial}{\partial
\theta}\right)
$$
Source terms $\epsilon$, $\sigma^{\phi}_{\phi}$,
$\sigma^{\theta}_{\theta}$, $\sigma^{r}_{r}$ are defined, according to the standard notations, as
\begin{equation}
\epsilon=\Gamma^2\left(\rho + \frac{p}{c^2}\right) -
\frac{p}{c^2},
\end{equation}
\begin{equation}
\sigma^{r}_{r}=\sigma^{\theta}_{\theta}=\frac{p}{c^2},\quad
\sigma^{\phi}_{\phi} = \frac{p}{c^2} + \left(\epsilon +
\frac{p}{c^2}\right) \frac{U^2}{c^2},
\end{equation}
\begin{equation}
p_{\phi}=e^{\mu}\left(\epsilon+\frac{p}{c^2}\right) \frac{U}{c} r
\sin\theta,
\end{equation}
where $\Gamma$ is the Lorentz factor
$$
\Gamma=\left(1-\frac{U^2}{c^2}\right)^{-1/2}, \quad
U=e^{\mu-\psi}(\Omega - \omega)r\sin\theta.
$$
and $U$ is the linear velocity of rotation. The equation for the scalar
curvature in quasi-isotropic coordinates has the following form:
\begin{equation}\label{EQ_R}
\triangle_{(3)}f'(R)=\frac{8\pi}{3}
e^{2\lambda}\left(\frac{3p}{c^2}-\rho\right)-\frac{e^{2\lambda}}{3}(f'(R)-2f(R))-\partial{
(\psi+\mu)}\partial{ f'(R)}.
\end{equation}
It is straightforward  to plug models $f(R)=R+\alpha R^2$ and $f(R)=R^{1+\varepsilon}$ into the system \eqref{EQ_N}-\eqref{EQ_R} and then to develop analysis for rotating case in analogy to non-rotating one.

\section{Numerical Results}
Considering the previous rotating and non-rotating cases, let us
report now results relevant to the conclusion we are looking for.
For a complete analysis of static stellar configurations see
\cite{Feola} and \cite{Capozziello2016}. 

\begin{figure}
  \includegraphics[scale=0.2]{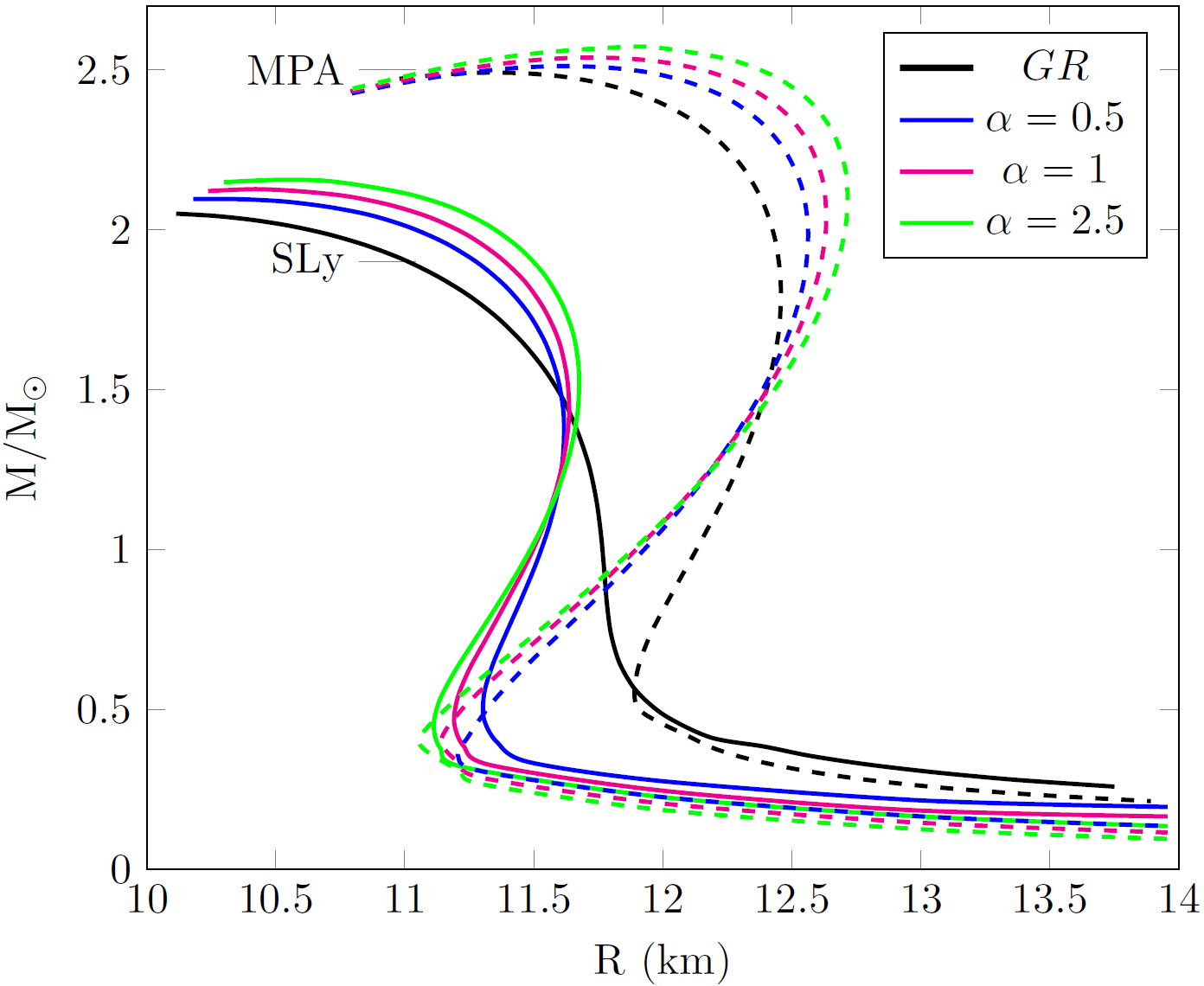}
  \caption{ $\mathcal{M-R}$ diagram for NSs in the $f(R)=R+\alpha R^2$ compared with GR considering the  SLy and MPA1 as EoS. Parameter $\alpha$ is given in units of $r_{g}^2=G^2 M_{\odot}^{2}/c^4$. Here $r_g$ is
  the gravitational radius of the Sun.}
  \label{fig1}
\end{figure}

For the $f(R)=R+\alpha
R^2$ model,   results are reported in Fig. \ref{fig1}. Here we
note that the larger the value $\alpha$ is, the larger the NS mass
becomes. It is immediate to see
 that, for appropriate values of $\alpha$,  we can  reproduce
the  values reported in \cite{Ligo}. Considering the MPA1 as  EoS,
 reported in  \cite{Feola},  the mass value of $2.6
M_{\odot}$ is easily achieved. This is more difficult considering the case SLy for EoS. See  Fig. \ref{fig1}.

It is interesting to consider both the influence of rotation with high
frequency and deviation from GR  on NS parameters. From the observational data, it follows that the highest measured
rotation frequency  is 716 Hz for the pulsar PSR J1748-2446ad
\cite{Hessels}. For various EoS, such a  frequency leads to an  increasing
of maximal mass of the order  $\sim 0.07\div 0.1 M_{\odot}$, in the GR context, which is not sufficient to explain the  data reported by LIGO \cite{Ligo}.

\begin{figure}
  \includegraphics[scale=0.2]{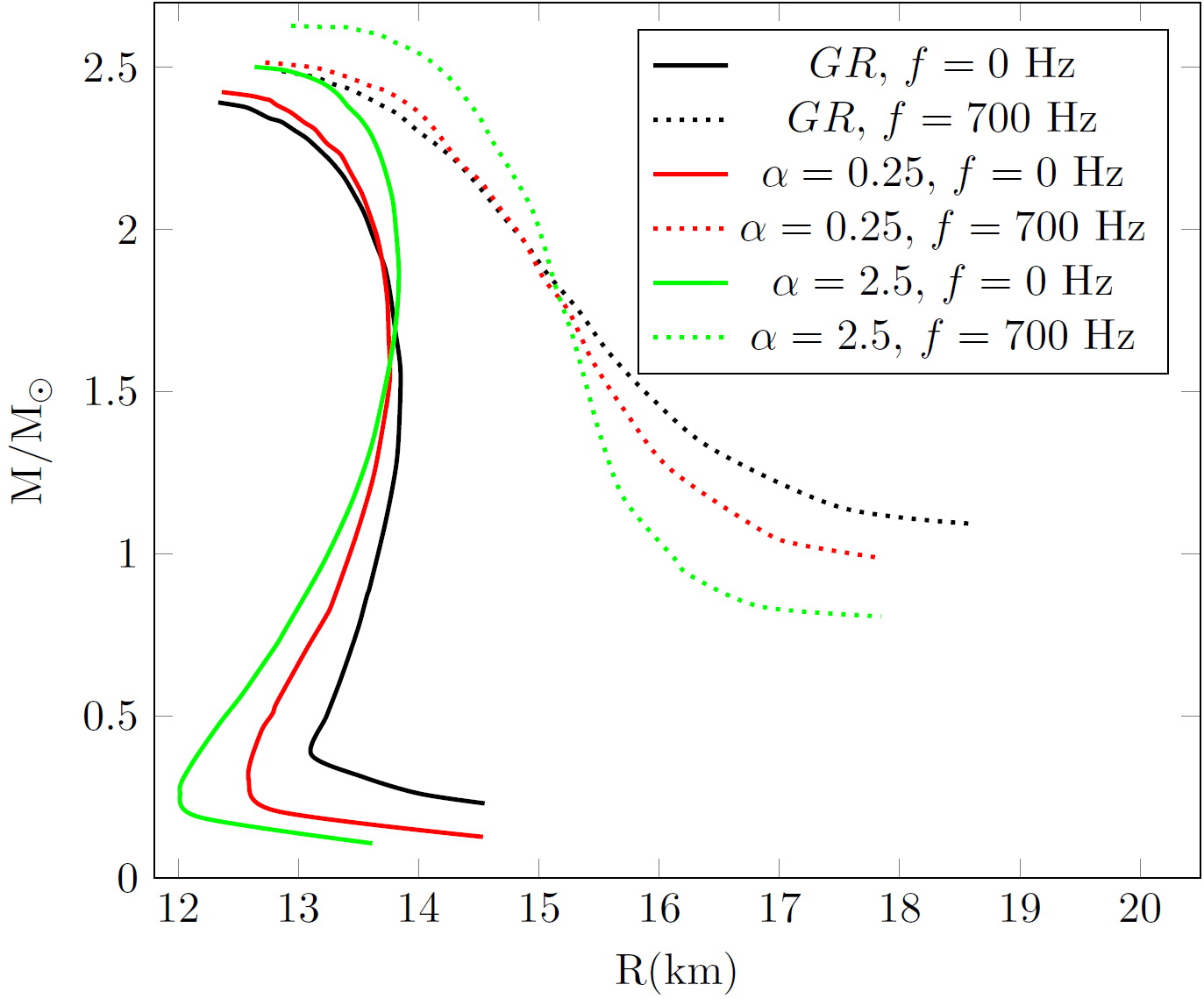}
  \caption{$\mathcal{M-R}$ diagram of NSs for $f(R)=R+\alpha R^2$  compared with GR. We are  considering the  GM1 as EoS without hyperons  in cases with and without  rotation.}
  \label{fig0}
\end{figure}

Let us consider, as an illustrative example, the EoS  GM1 without
hyperons \cite{GM}  and, for frequency,  let us  assume the value
$f=700$ Hz. Results of our calculations show that maximal mass for
non-rotating stars in GR is $2.39 M_{\odot}$. For stars rotating
with $f=700$ Hz, the value increases up to $2.49 M_\odot$. For
$f(R)=R+\alpha R^2$ gravity, the maximal mass of static star is
$2.50 M_\odot$ assuming $\alpha=2.5$. In the case of rotation with
$f=700$ Hz, the maximal mass is $2.63 M_\odot$ showing that the
LIGO limit can easily  be  achieved  (see Fig. \ref{fig0}).

According to the data in \cite{Cipolletta}, the maximal mass in the case of uniform
rotation for GM1 as EoS is $2.84 M_\odot$ assuming  GR. However,
this  mass-shedding limit is reached for a Keplerian frequency of 1.49 kHz and the existence of  so fast rotating stars seems unrealistic. On the other hand,   it seems  possible that, in the context of $R+\alpha R^2$ gravity,  supermassive NSs,  with masses close to $3M_\odot$, can appear
for observed rotation frequencies.

Furthermore, it is worth  mentioning that some stiff EoS were proposed with the
 maximal mass limit  for non-rotating stars in the range $\sim 2.75\div 2.8\,
M_\odot$ (see for example MS1 \cite{MS1}, NL3 \cite{NL3}). For
frequencies $\sim 700$ Hz, in $R+\alpha R^2$ gravity with large
values of $\alpha$, the maximal NS mass can also be  close  to
$3M_\odot$.

\begin{figure}
  \includegraphics[angle=-90,scale=0.3]{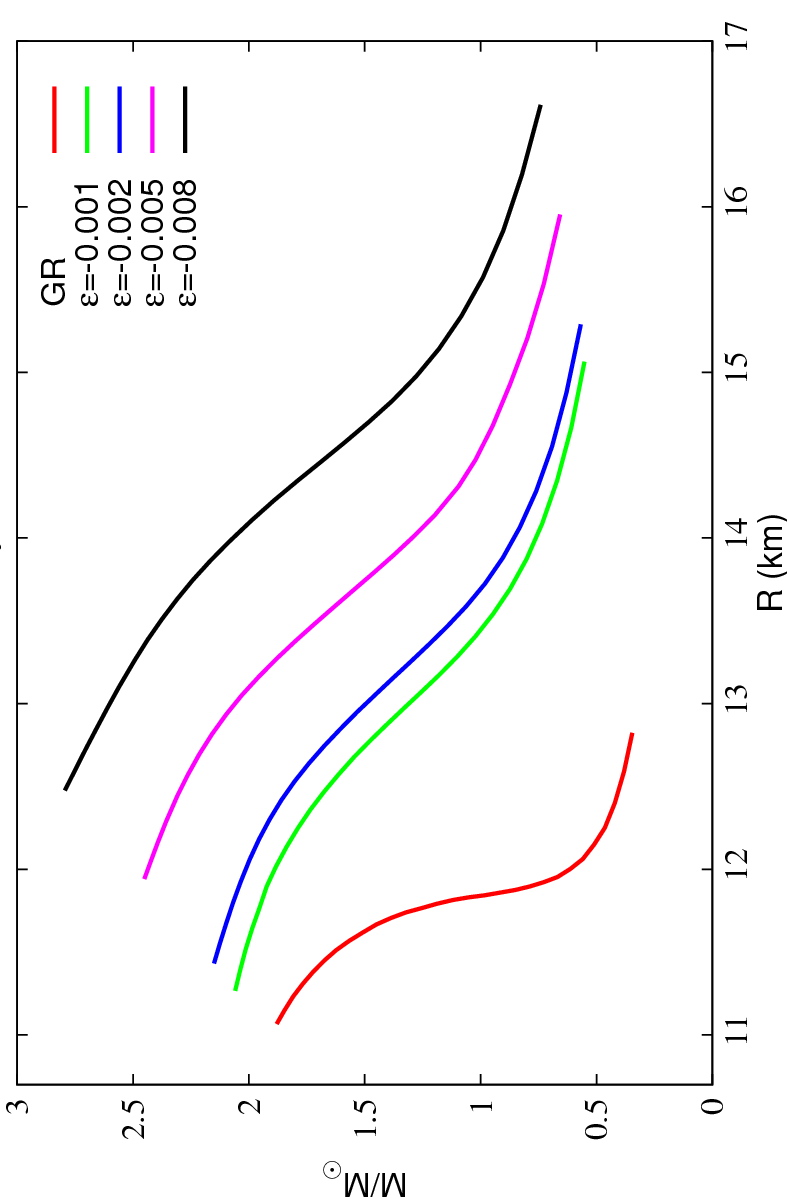}
  \caption{ $\mathcal{M-R}$ diagram for NSs in the $f(R)=R^{1+\varepsilon}$ model considering the SLy as  EoS. The relation between the parameter $\varepsilon$ and the mass $\cal M$ is evident.}
  \label{fig3}
\end{figure}

In the case of $f(R)=R^{1+\varepsilon}$, following \cite{Feola,
Capozziello2016}, we adopt SLy as EoS and the results of our
numerical analysis are shown in Fig. \ref{fig3}. Here we can
notice that the value of $\varepsilon$ influences greatly the
$\mathcal{M-R}$ relation. In particular the larger $\varepsilon$
is, the larger the the NS mass becomes. In Fig. \ref{fig3}, we
reported the $\mathcal{M-R}$ relation for $\varepsilon$ between
0.005 and 0.008, which are consistent with the mass in the range $
2.50\div 2.67\,M_{\odot}$ reported by \cite{Ligo}.

\section{Conclusions}

In this paper, we  presented a way to  theoretically  explain the
anomalous mass of  compact object  recently detected by
\cite{Ligo} with the hypothesis that it is a supermassive  NS.
Specifically,  for $f(R)=R+\alpha R^2$ gravity with maximal
observed rotation and for $f(R)=R^{1+\varepsilon}$ gravity without
rotation, it is straightforward to obtain results consistent with
LIGO detection without invoking exotic EoS. The fact that ETGs are
consistent with observations which cannot be explained by standard
GR is fundamental not only because we can shed new  light on the
extreme gravity regimes that are realized in compact objects like
NSs, but also because these observations could validate more and
more the theoretical grounding of ETGs. It is worth noticing that
it could be not only an alternative explanation of the reported
results, but a sort of  {\it experimentum crucis} for these
theories, if such a kind of  (present or future) observations cannot
be explained in the framework of GR.

\section*{Acknowledgements}
S.C.  acknowledges the support of INFN (iniziative specifiche MoonLIGHT-2 and QGSKY).

\end{document}